\documentclass[aps,prl,preprint,superscriptaddress, 11pt]{revtex4-1}
\usepackage{xcolor}						
\usepackage{graphicx}	
\usepackage{amssymb}
\usepackage{amsmath}
\usepackage{float}
\usepackage[caption = false]{subfig}
\usepackage{natbib}

\def\beq{\begin{equation}}
\def\eeq{\end{equation}}
\def\rmd{{\rm d}}

\graphicspath{{./}{figures/}}

\begin{document}

\title{On the principle of Astrometric Gravitational Wave Antenna}
\author{Mariateresa Crosta}
\email[]{mariateresa.crosta@inaf.it}
\affiliation{Astrophysical Observatory of Turin (OATo) - INAF, via Osservatorio 20, I-10025 Pino Torinese (TO), Italy}
\author{Mario G. Lattanzi}
\affiliation{Astrophysical Observatory of Turin (OATo) - INAF,  via Osservatorio 20, I-10025 Pino Torinese (TO), Italy}
\author{ Christophe Le Poncin-Lafitte}
\affiliation{SYRTE, Observatoire de Paris, Université PSL, CNRS, Sorbonne Université, LNE, 61 avenue de l’Observatoire, 75014 Paris, France }
\author{Mario Gai }
\affiliation{Astrophysical Observatory of Turin (OATo) - INAF, via Osservatorio 20, I-10025 Pino Torinese (TO), Italy}
\author{Qi Zhaoxiang}
\affiliation{Shanghai Astronomical Observatory, CAS 80 Nandan Road, Shanghai 200030, China}
\author{Alberto Vecchiato}
\affiliation{Astrophysical Observatory of Turin (OATo) - INAF, via Osservatorio 20, I-10025 Pino Torinese (TO), Italy}

\begin{abstract}
 
The direct detection of gravitational waves by ground-based optical interferometers has opened a new window in astronomy. Nevertheless, as these detectors are a combination of two Michelson-Morley like baselines, their sensitivity for determining the incident direction of a gravitational wave is quite weak compared to current high-precision of space astrometry. 
We therefore present a novel concept for a gravitational wave antenna in space that uses close pairs of point-like sources as natural sensors to record and characterize the very tiny variations in angular separations induced by a passing gravitational wave, thus operating  complementarily to linear arm detectors and enabling to gain informations on the gravitational wave incoming direction. 
Indeed, the proposed astrometric gravitational wave observable builds on methods of relativistic astrometry that can substantially enhance the strength of gravitational wave signals by exploiting to the fullest the telescope optical resolution and, at same time, provide a powerful tool for identifying the direction to the sources that originated the gravitational wave by monitoring close pairs of stars. Furthermore, the use of two local-line-of-sights in a differential  formulation avoids issues related to high order modeling of the local (Solar System) background geometry and the need for accurate satellite ephemeris and attitude.  
 


\end{abstract}

\maketitle




\section{Introduction}

Experimental confirmation of gravitational waves (GWs) through the LIGO \citep{ligo, adligo} and VIRGO \citep{virgo} antennas have gained  great impulse to the search and characterization of  possible GW sources. New ground-based experiments are joining \citep{kagra} or in the making \citep{ET, CE, decigo} and the LISA mission \citep{LISA} will soon implement similar concepts in space. The prime objective is the complete description of GWs, i.e., to determine their amplitude and frequency, and then pinpoint their directions for multi-waveleng “profiling”. 
A recent review on gravitational-wave physics and astronomy in the present decade can be found in \citet{roadmap}.

Astrometric observations are primarily targeted for the accurate determination of directions to the incoming photons and their change with time. Therefore they collect photons that have interacted with different time-dependent gravitational fields along their propagation. With the advent of highly accurate astrometric (angular) and radial velocity measurements in space, observation reduction models compliant with General Relativity (GR) have become a necessity. This is because, when compared to the targeted measurement precision levels, the amount of  space-time curvature due to the Sun and to all of the other relevant Solar System (SS) masses (including the Earth-Moon system) that are affecting incoming photons and space ``observers'' alike can no longer be ignored. 
Weak gravity regime influences electromagnetic propagation on a much wider domain than its strong counterpart.\\
A gravitational wave detection via astrometry was explored by several authors \citep{braginsky, payne, kaiser, gwinn1997, book, mihaylov2018}. In practice, the main issue is   to consider deflection extra-shifts induced by passing GWs on photons propagating to an observer from within the SS. Such a detection is not considered promising, since it requires, at lower order, the nanoarcsecond level of accuracy \citep{damour, crosta1998, shutz}  and depends on the knowledge of the satellite attitude. So far, preliminary analysis has been focused on periodic GW signals with period shorter than the Gaia operating time \citep{moore2017, klionercqg} or on secular effects \citep{klomig} in QSO proper motions for longer period, i.e. for  ultra-low GW frequencies. \\ 
As a matter of fact, the Gaia-like astrometric observable is the direction cosine between incoming stellar light and the observer attitude-tetrad \citep{crosta2017}. Its strict application to a passing GW generates serious shortcomings. The addition of a correction to the direction cosine solely due to the passing GW signal, ignoring the instrumental contribution of the satellite uncertain attitude, would be not the optimal strategy for our purpose, i.e. to find the incoming GW signal, as it would impose a requirement on the accuracy of the attitude knowledge to implement in-orbit. 
Besides the above, disentagling the GW signals requires that the SS background metric should be consistently developed to account for possible background (natural) ``noises'', thus preventing systematic effects in modelling our astrometric observable. Hence, since $v/c \sim 10^{-4}$ rad, being $v$ the typical velocity of each relevant SS metric source, one should retain terms in the null geodesic solution  up to $(v/c)^4$ at best, i.e., at the nanoarcsecond level (nas). \\
Here we present a novel idea of an $\it{Astrometric~ GW~ Antenna}$ 
that uses angles, i.e. differential astrometry,  instead of linear baselines as the current “spatial” antennas taking advantage from the relativistic astrometric techniques successfully developed for Gaia \citep[ESA,][]{gaia, crosta2017}.

%

To overcome the limitations inherent in the measurement of a single light direction, in the following we introduce a different fundamental observation equation for the astrometric GW detection based on the angle relative to pairs of local line-of-sights. Finally, we also give a first evaluation of its potential impact on GW science and its practical feasibility, in the light of new, specialized and much improved concepts for space astrometry missions \citep{sait, astra1, astra2}.


\section{The fundamental observation equation for the space-born astrometric gravitational wave antenna}

Given the global metric due to both SS sources and passing/standing GW perturbations at the observer's location in the form 
\begin{equation}
{ g}_{\alpha \beta}=  {g}^{SS}_{\alpha \beta} + {h}_{\alpha \beta}^{GW} = \eta_{\alpha \beta} +  \sum_{(a)} {h}_{(a) \alpha \beta}^{SS}+  {h}_{\alpha \beta}^{GW} + O(h^2) ~ ,
\label{globalmetric}
\end{equation}
where $ \eta_{\alpha \beta} $ is the flat Minkowskian metric, with signature (-,+, +,+), the  cosine of the angle between two observed light directions ${\bar \ell}^{\alpha}_1$ and  ${\bar \ell}^{\alpha}_2$ writes (see appendix A for the derivation)
\begin{equation}
\cos \psi_{1,2} = {g}_{\alpha \beta}  ({\bar \ell}^{\alpha}_1 {\bar \ell}^{\beta}_2 )_{obs}~,
\label{directioncosine}
\end{equation}
 where  ${\bar \ell}^{\alpha}$ is the null tangent unit four-vector projected on the rest space  of the local barycentric observer, namely, for of our solar system, the observer at rest relative to the barycentric celestial reference system (BCRS)  with coordinates $(t,x^i)$  \citep[and references therein]{crosta2019}.  The differential nature of the measurement greatly relaxes requirements on a precise satellite attitude and payload (thermal and mechanical) stability. 
 


Similarly, stellar light directions can be separated into the SS part (due to the background metric) plus a perturbation shift, i.e. $\delta \ell^{\alpha}$, attributed purely  to the passing GW:
\begin{equation}
{\bar {\ell}}^{\alpha}_{obs} =  \bar \ell^{\alpha (SS)} + \delta \ell^{\alpha (GW)} + O( \delta \ell^2).
\label{lightdirection}
\end{equation}
Then, the cosine can be further simplified if $\delta \ell^{(GW)}\sim \epsilon^4 $ as follows
\begin{equation}
\cos \psi_{1,2} =\cos \psi_{1,2}^{SS} + \eta_{\alpha \beta} ({\bar \ell}^{\alpha}_{1_0} \delta \ell_{2}^{\beta(GW)} +  {\bar \ell}^{\alpha}_{2_{0}} \delta \ell_1^{\beta (GW)})_{obs}+ h^{GW}_{\alpha \beta} {\bar \ell}^{\alpha}_{1_0} {\bar \ell}^{\beta}_{2_{0}} + O(\epsilon^5)+ O(h^2),
\label{eq:alternativy}
\end{equation}
 where ${\bar \ell}^{\alpha}_{i_0}$ represents the unpertubed light direction to star i=1,2. 

 It is straightforward to check that the cosine of the total angle,  cannot be simply split into a background direction cosine plus its GW analogue
 (see appendix B for details).
 On the other hand, since the passing GW produces an extra shift on the light deflection, we expect a perturbation $\delta \psi^{GW}$ to the undisturbed angle between two light directions.
 \\
Then, we can expand our observation equation as follows:
\begin{equation}
\cos (\psi_{1,2}^{SS}  + \delta \psi_{1,2}^{GW}) =  \cos(\psi_{1,2}^{SS}) \cos( \delta \psi_{1,2}^{GW}) -   \sin(\psi_{1,2}^{SS}) \sin( \delta \psi_{1,2}^{GW})=     \cos \psi_{1,2}^{SS} + F^{GW}_{1,2},
\end{equation}

where $F^{GW}_{1,2}\equiv  \eta_{\alpha \beta} ({\bar \ell}^{\alpha}_{1_0} \delta \ell_{2}^{\beta (GW)} +  {\bar \ell}^{\alpha}_{2_0} \delta \ell_1^{\beta (GW)})_{obs}+ h^{GW}_{\alpha \beta} {\bar \ell}^{\alpha}_{1_0} {\bar \ell}^{\beta}_{2_0}  + O(\epsilon^5)+ O(h^2)$.

Since $\delta \psi^{GW}\ll 1$, the previous equation reduces to consider the unknown angle due to the passing GW, i.e. 
\begin{equation}
\delta \psi_{1,2}^{GW} =  - \frac{F^{GW}_{1,2}}{\sin(\psi_{1,2}^{SS})}.
  \label{obseqper}
\end{equation}

%
%
%
The new  expression (\ref{obseqper}) deduced from observation equation (\ref{directioncosine}) depends on the observed perturbation to the $\psi_{1,2}^{SS}$  angle caused by the passing GW in the $F^{GW}_{1,2}$ term; the observability of this angular perturbation can be boosted through the factor (sin$ (\psi_{1,2}^{SS}))^{-1}$ that acts as a ``signal amplifier'' for the GW detection, being mainly limited by the resolving power of an astrometric telescope that  never  goes to zero because of physics. \\
 As we are putting forth a novel operational principle for measuring GWs that takes great advantage from the quantity $\psi^{SS}_{1,2}$, here it will suffice to indicate how that can be estimated/derived from actual measurements of the pair separation. What we measure is actually the angle $\psi_{1,2}(t_i)$ between point-like sources '1' and '2' at time $t_i$. The measurements $\psi_{1,2}(t_i)$  are taken with a high cadence, i.e., with frequency  $\omega_{S} \gg  \omega_{GW}$, the oscillating frequency of the GW, assumed 'monochromatic'. This last condition ensures that the Nyquist-Shannon (sampling) theorem is satisfied and, at the same time, sufficient statistics is built to beat (single) measurement noise. \\
Let's then take the average over N ($\gg$ 1), i.e. the number of samples of the pair angular distance taken over a session, of the measurements $\psi_{1,2}(t_i)$. We have:
$ < \psi_{1,2}(t_i)>_N$ $= < \psi^{SS}_{1,2} >_N + < \delta \psi^{GW}_{1,2} (t_i) >_N$ $\simeq \hat{\psi}^{SS}_{1,2}$, as we can make the average $< \delta \psi^{GW}_{1,2} (t_i) >_N$ as small as needed. 


With the evaluation of the unperturbed angular separation of our antenna arm (the point-like pair) directly from the observations, as shown above
, it follows that: $ \delta \hat{\psi}^{GW}_{1,2} (t_i) \equiv \psi_{1,2}(t_i) - \hat{\psi}^{SS}_{1,2}$.
This is the GW perturbation to the antenna angle as estimated directly from the measurements that can be actually used, together with $\hat{\psi}^{SS}_{1,2}$, in Eq.(\ref{obseqper}) to build the observation equations from which strength and direction to the GW source can be derived.  
The measurement and initial data processing strategy/protocol just sketched supports the theoretical practicability of our concept for an astrometric GW antenna.   
\\
 Equation (\ref{obseqper}) requires to define (i) the $h^{GW}$ at the observer, (ii) a pair of unperturbed local line-of-sights, $ \bar \ell^{\alpha}_{obs}$, from the geodesic equation with the metric (\ref{globalmetric}), and (iii) the corresponding shift due to the passing GW.
 
 The general form of the GW perturbation can be expressed  as a function of argument $\tilde k_{\alpha} x^{\alpha}$, namely 
$h_{ij}^{GW}(\tilde k_{\alpha} x^{\alpha})$ 
with tangent vector  $\tilde k^{ \alpha} =  \tilde k^0 \partial_0^{ \alpha} +  \tilde k^i \partial_i^{ \alpha},$
where  $p^i = \tilde k^i/ \tilde k^0$ 
is the direction  of the GW propagation.

 Finally, in the linearized regime, the gravitational wave shift is recovered within the suitable astrometric models, e.g., those adopted for the data analysis and processing of the Gaia satellite \citep[and reference therein]{crosta2017}, via the geodesic integration (see appendix C), namely:
\begin{equation}
\int^{\sigma_*}_{\sigma_{obs}}  d \delta \ell^i =  \frac{\bar \ell^i_{_0} +  p^i}{2 (1 + p \cdot \bar\ell_{_0})}  \bar \ell^j_{_0} \bar\ell^k_{_0} \int^{\sigma_*}_{\sigma_{obs}}   d h_{jk}^{(GW)}  (\sigma)  -   \frac{\bar\ell^j_{_0}} {2}   \int^{\sigma_*}_{\sigma_{obs}} d h_{ij}^{(GW)}  (\sigma) + O(\epsilon^5),
\end{equation}
that in the far away zone reduces to
\begin{equation}
\delta \ell^i =  \frac{ \bar \ell^i_{_{0}} + p^i}{2 (1 + p \cdot \bar\ell_{_0})}  \bar\ell^j_{_{0}} \bar\ell^k_{_{0}}  h_{jk (obs)}^{(GW)}  -   \frac{\bar \ell^j_{_{0}}} {2}   h_{ij (obs)}^{(GW)}  + O(\epsilon^5),
\end{equation}
explicitly showing that $\delta \ell^i \propto h^{GW}$, and in agreement  with what found in \citet{payne} and \citet{book}. In such a case,  our expression (\ref{obseqper})  then becomes

\begin{eqnarray}
 \delta \psi_{1,2}^{GW}  &= &   -  \frac{ h^{GW}_{ij (obs)}}{2 \sin(\psi_{1,2}^{SS}) }  \left\{  \frac {[(\bar \ell_{1_0} \cdot \bar \ell_{2_0}) +  (\bar \ell_{1_0} \cdot  p) ]  \bar \ell^i_{ 2_0}\bar \ell^j_{ 2_0}}{(1 + p \cdot \bar \ell_{2_0})} +   \frac { [(\bar \ell_{2_0} \cdot \bar \ell_{1_0}) +  (\bar \ell_{2_0} \cdot  p) ] \bar \ell^i_{ 1_0}\bar \ell^j_{ 1_0}}{(1 +  p \cdot \bar\ell_{1_0})} \right\} \nonumber \\
 && +  O(\epsilon^5)
 \label{eqfinal}
 \end{eqnarray}
 
  which shows also the dependence on the scalar products between the SS direction of propagation of photons from stars $1,2$ (i.e., alternatively, the angle $\psi_{1_0, 2_0}$) and that of  each star direction with respect to the GW source (angles $\psi_{1_0, p}$ or $\psi_{2_0, p}$).

\section{The operating principle of the GW astrometric antenna}

\begin{figure}[htbp]
\centering
\subfloat[]{\includegraphics[width=.55\textwidth]{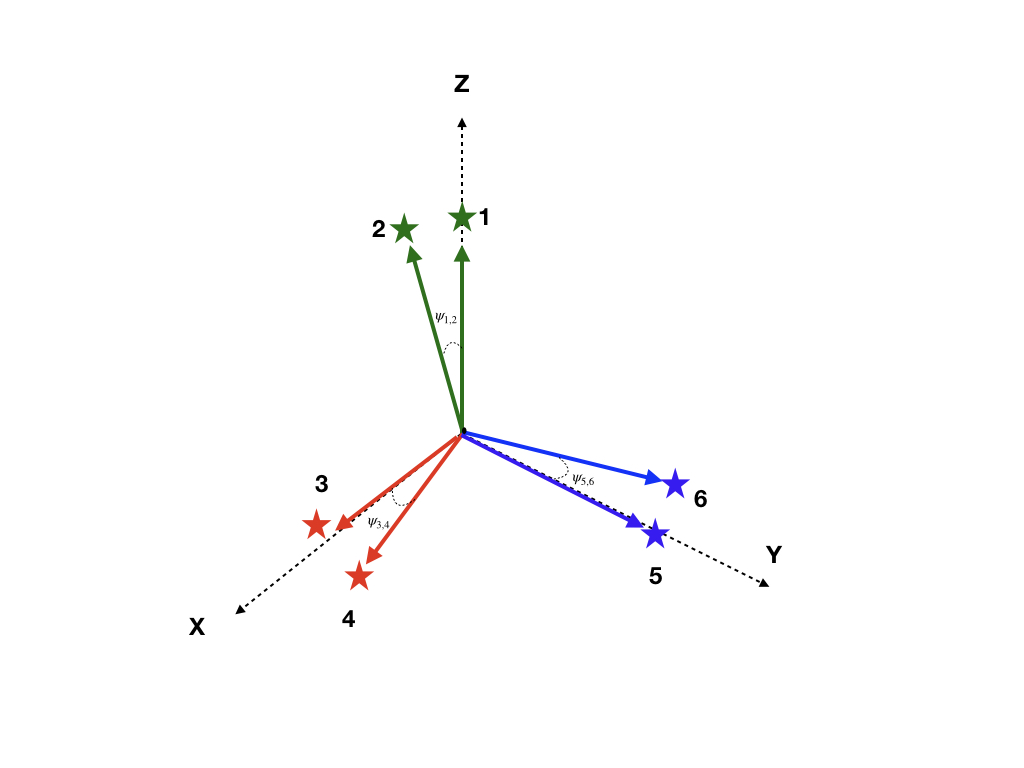}} \\
\subfloat[]{\includegraphics[width=.50\textwidth]{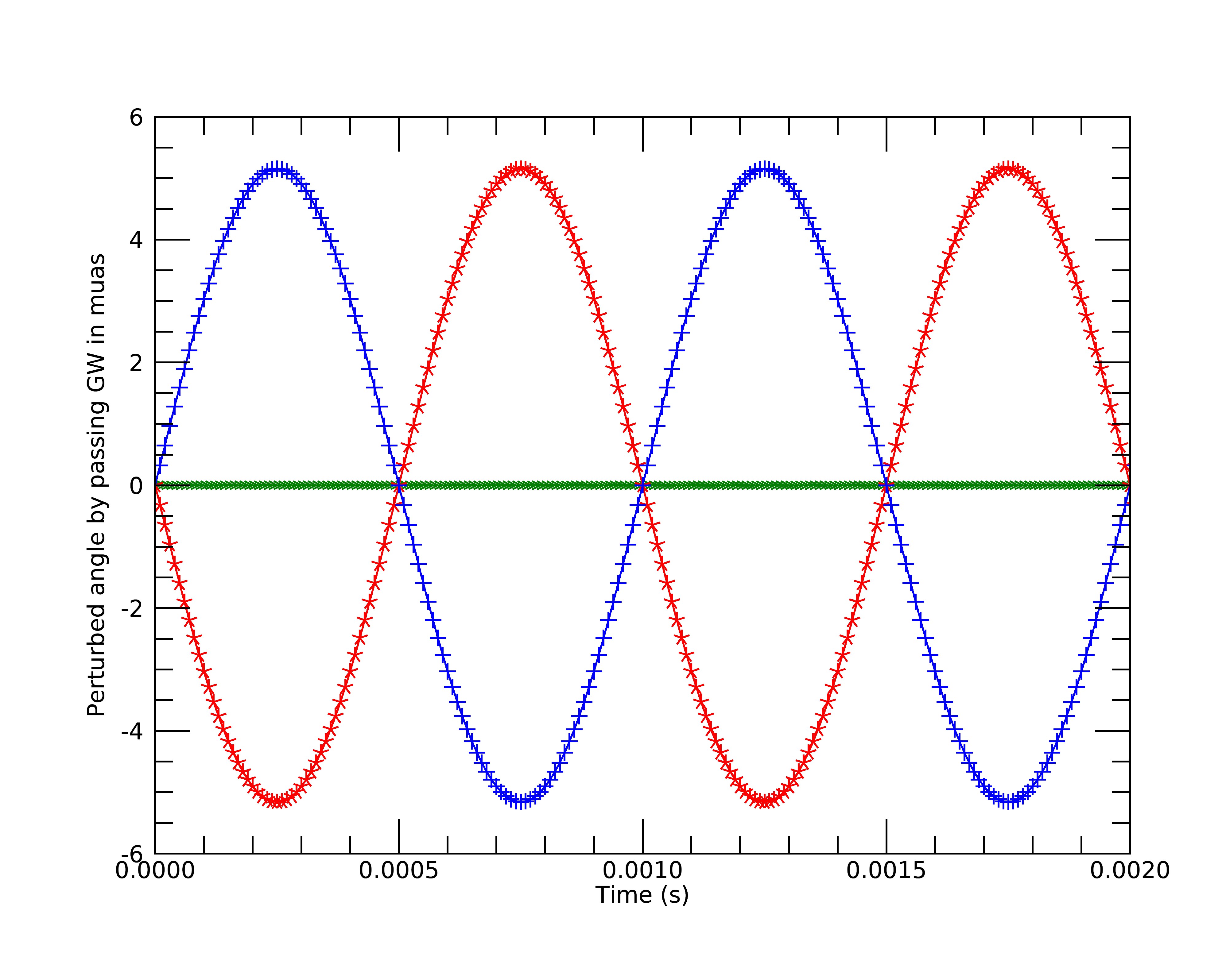}}
\subfloat[]{\includegraphics[width=.50\textwidth]{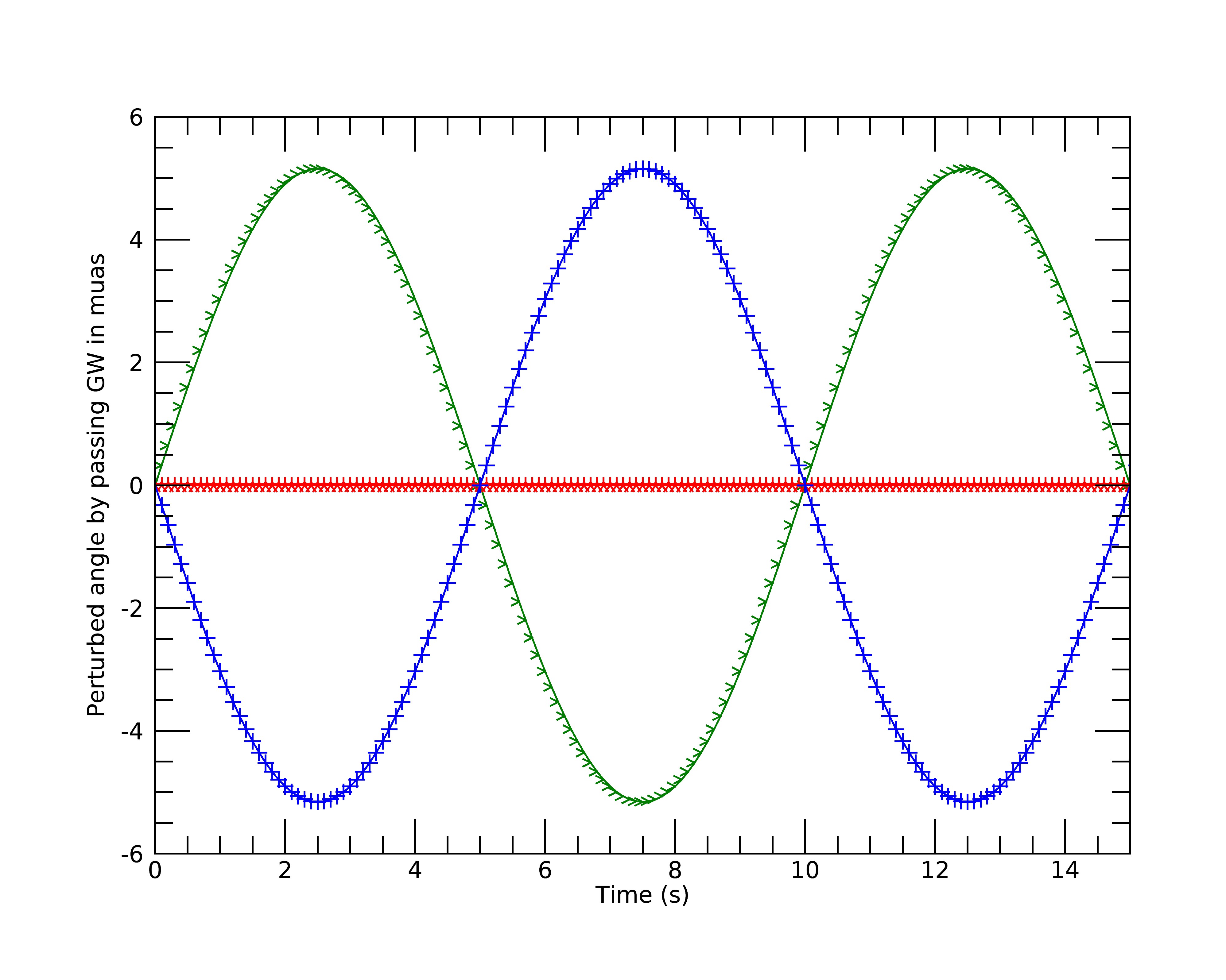}}
\caption{In (a) a possible configuration of a 3-LOS telescope, the ($\mathbf{x}$,$\mathbf{y}$, $\mathbf{z}$) triad, with respect to the chosen directions:   the local (SS) vectors  $\bar \ell_{3_0}$, $\bar \ell_{5_0}$ and $\bar \ell_{1_0}$ to stars 3,5 and 1, respectively. The angles ${\psi}_{i, j}$, representing the instantaneous angular distances of the stellar pairs 
are all assumed small at $\sim 0.01''$. With the same colour coding, the other two panels show, fixed the resolution limit, the astrometric signals $\delta\psi^{GW}_{i,j}$ (from Eq. (\ref{eqfinal})) caused by a GW passing along: (b) the $\mathbf{z}$ direction (green line) with  $A_{+}\sim 10^{-18}$ and frequency  $10^3 Hz$; (c) the $\mathbf{x}$ direction (red line) with amplitude $A_{+}\sim 10^{-19}$ and frequency 0.1 $Hz$.
The other curves indicate the GW-induced signals along the other two LOS's of the astrometric antenna.}
\label{fig:plot}
\end{figure}

 Equation (\ref{eqfinal}) above shows the operating principle of the astrometric antenna, and clearly  demostrates its direct relation with the direction to an incoming GW and, therefore, the potential to pinpoint its source.  For, the GW term  $h_{ij}$ (and its time variation) will mostly characterize the detection (the amplitude and phase term), while the factor within the  curly brackets 
will assume specific patterns according to the direction of the incoming GW relative  to the observer (spatial) orientation, i.e.  the ${\mathbf x}, {\mathbf y}, {\mathbf z}$ triad 
depicted in Fig. \ref{fig:plot} (a).  
 In the “Transverse and Traceless” (TT) standard gauge  the GW components are 
$h_{0 i}=0$,  $\delta^{ij} h_{ij}=0$,  $\delta^{ij} h_{jk,i}=0$, with only two independent degrees of freedom, the two amplitudes $A_{+}$ and $A_{\times}$.

Taking advantage of the property that in the TT gauge only the components perpendicular to the direction of propagation survive, below we show, without loosing too much to generality, how a GW reaching a space-born observer from within the SS can be astrometrically measured.\\
\noindent
Therefore, to provide quantitative examples suggestive of the potential of Eq. (\ref{eqfinal}), it will suffice to consider a plane GW, 
 of a given frequency $\omega$, linearly polarized (i.e., $A_{\times}=0 $) 
 , traveling in the direction $ \vec p$ assumed aligned with the direction $\bar \ell_{ 1_0}$, i.e.,  the local (SS) unperturbed direction to star 1.  
This is the same direction at which line-of-sight (LOS)  ${\mathbf z}$, of a three-LOS telescope materializing the ``observer'' of Fig. \ref{fig:plot}(a), is pointing. 
Also, $\bar \ell_{ 2_0}$, the local direction  to star 2, is at a very small angular separation to $\bar \ell_{ 1_0}$, i.e., $\cos (\psi_{1_0,2_0}) \sim 1$. 
Similarly to the LOS in the ${\mathbf z}$ direction of our schematic 3-LOS (or 3-way) telescope, one can align the other two viewing directions, ${\mathbf x}$ and ${\mathbf y} $, along the local directions  $\bar \ell_{3_0}$ and $\bar \ell_{5_0}$ to stars 3 and 5, respectively (Fig. \ref{fig:plot} (a)), while directions $\bar \ell_{4_0}$ and $\bar \ell_{6_0}$, to stars 4 and 6, are the corresponding close optical companions. Close stellar (or stellar-like) pairs here means that we are always concerned  with angles ${\psi}_{i_0, j_0}$ $\lesssim$  0.01 arcsecond, a number representative of the operational  resolution limit reached with telescopes already operating (e.g. HST) or 
 that will soon operate in space (like, e.g., NASA's NextGen Space Telescope, ESA's Euclid or CNSA's CSST, see further below for more) at optical wavelengths, i.e. $\geq 550$ $nm$, including the near-IR (to $\sim 2$ micron). 


\begin{table}[htbp]
   \centering
   \begin{tabular}{c | c | c | c |c  } 
      \toprule
$A_{+}$ &  $\psi_{i_0,j_0}$ &$max(\delta\psi^{GW}_{1,2})$  &  $max(\delta\psi^{GW}_{3,4})$ &  $max(\delta\psi^{GW}_{5,6})$    \\
(radians)   & (")   & ($\mu$as) & ($\mu$as) & ($\mu$as)\\ 
\hline

$10^{-18}$ & 0.01& 4.12$ \times 10^{-15} $     & 5.12    &  5.12    \\
$ 10^{-19}$ & 0.01 & 4.12$ \times 10^{-16} $ & 0.51 &   0.52 \\
$10^{-20}$& 0.01&  4.12$ \times 10^{-17}$   & 5.12 $\times 10^{-2}$ &  $5.12 \times 10^{-2}$   \\
$ 10^{-21}$ & 0.01 & 0  & $ 5.12 \times 10^{-3} $  &  $ 5.12 \times 10^{-3} $   \\
\hline
$10^{-18}$  & 0.001 & 4.12 $ \times 10^{-16} $ & 51.57 &  51.57  \\
\hline
\hline
   \end{tabular}
   \caption{Angular perturbations, $max(\delta \psi^{GW}_{i,j})$ ($\mu$as), for different linear strains of amplitude $A_+$ (radians) propagating along ${\mathbf z}$. The values derive from Eq. (\ref{eqfinal}) after setting the three antenna arms $\psi_{i_0,j_0}$ to the values in the second column (in arcsec). As expected, decreasing $\psi_{i_0,j_0}$  by one order of magnitude, the corresponding $\delta \psi^{GW}_{i,j}$ is amplified by the same amount. This is the case for the values reported in the last row compared to its direct analogue in first. Note that along the GW direction of propagation $\delta \psi^{GW}_{1,2}$ is not null, as the direction to star 2, forming the 1-2 pair, although quite small, is not coincident with the LOS to star 1.}
   \label{tab1}
\end{table}

We are now able to provide actual examples utilizing representative cases of GW strains known from the literature. The first column of Table \ref{tab1} presents the amplitude, $A_{+}$, of metric perturbations from possible GW sources as described in \citet{thorne87} and \citet{ColpiSesana}. The last three columns provide, for each $A_{+}$, the maximum (angular) perturbations resulting on the local unperturbed (angular) separations of the pairs (the "arms" of the astrometric antenna) along the three telescope axes (Fig. \ref{fig:plot} (a)). The perturbation signals can be calculated from Eq.s (\ref{eqfinal}) 
for a plane GW traveling along the ${\mathbf z}$ direction and ${\psi}_{i_0, j_0}$ set to the values in Table I. Gravitational strains as large as $h \sim 10^{-18}$ are associated with SN core collapse events, thus, they are short lived and at frequencies $\omega \sim$ 10$^3 Hz$. To the range of high frequency sources, $\omega$=[10 -- 1,000] $Hz$, belong also the cases of coalescing compact binary systems (NS-NS, pairs of stellar black holes, BH$^*$-BH$^*$, or NS-BH$^*$). 
Besides, at $\omega$ $<$ 1 $Hz$ and in a range of characteristic strain amplitudes spanning 3 orders of magnitude, from $\sim$ 10$^{-18}$ to 10$^{-21}$, one finds not only coalescing super-massive BH's, but also the significantly more numerous population of resolved and unresolved Milky Way binaries, with at least one degenerate companion. 
Figure \ref{fig:plot} (b) shows 5 $ms$  (i.e. 5 times the simulated period) of the angular perturbation $\delta\psi^{GW}_{i,j}$ experienced by the three ''arms''  (the angles ${\psi}_{i_0, j_0}$) under the strain of the high frequency-high amplitude  case mentioned before.
As expected, with a GW propagating in the direction of the ${\mathbf z}$-axis, the 1-2 stellar 
pair is practically unperturbed, while all the action is with the pairs along the 
$\mathbf{x}$ and $\mathbf{y}$ axes.

If statistically meaningful measurements of variation of angular separations 
of $\sim$ 5 $\mu as$, as in Fig. \ref{fig:plot} (b), although possible,
are yet to be proven, variations 10 times larger 
are already within the capabilities of present-day, or soon-to-be, space astrometry missions, like those mentioned earlier (see \cite{sait} and references therein). 
This is possible also because actual (angular) resolution limits, governing the amplifications of the perturbation signal associated with the GW (Eq.(\ref{eqfinal})), can be made significantly higher by utilizing optimal calibrations procedures 
\citep{bernacca1,bernacca2, steffen}.
Signal frequencies of 1 $kHz$, as in Fig. \ref{fig:plot} (b) are hard
to achieve with high signal-to-noise ratio, as sampling above the Nyquist-Shannon frequency would call for integration times shorter than 0.5 $ms$. 
Typical ''fast'' exposure times are at 25 $ms$ in the case 
of the astrometer FGS aboard HST \citep[and references therein]{bucciarelli}, and only 100 $ms$ on Euclid \citep[and references therein]{bosco}, clearly not enough. And, simply allowing for much shorter exposure times might  not improve things as, given the size of the telescope mirrors on these missions, it might become difficult to reach sufficiently faint magnitudes,  thus forcing the 3-LOS telescope to limit its orientations to directions with bright stellar pairs. 

Besides, as mentioned earlier, GW sources with similar amplitudes, or even larger, are expected at more comfortable frequencies, making this scenario a clear case for GW science with an astrometric antenna.  
Figure \ref{fig:plot}(c) illustrates the progress of $\delta\psi^{GW}_{i,j}$ with time for the GW strain amplitude in row 2 of Table \ref{tab1}; $\omega =$  0.1 $Hz$,  while the separations of the angular arms are here set to 0.001$"$, instead of 0.01$"$, thus generating variations of a few $\mu$as. Therefore, this figure and the  considerations on the cases presented in Table \ref{tab1}, confirm  that an astrometric antenna, capable of monitoring periodic signals of amplitude $>$ 1 $\mu$as, possesses the ability of measuring GW's associated with a range of coalescing massive BH's events.

Finally, row 4 of Table I refers to smaller amplitudes as expected for GW's from core collapse events (at $\omega \sim 100 Hz$), or, at much shorter frequencies, from resolved and unresolved MW binaries. These  cases appear beyond today's  technology on current missions, including payloads that will be flying into orbit within the next few years.

Expanding the access to the physics of GW's for an astrometric antenna would then require dedicated technological developments 
\citep[see e.g. ][]{riva10, sait}.
Regardless of the actual detectability the priority of this article is to prove the concept of a novel idea that might develop 
into new space-borne astrometric instrumentation for the study of GW's. From this perspective, it is evident that the angular arms of the 3-LOS GW antenna, as schematically depicted in Fig. \ref{fig:plot} (a), would register the astrometric signals with different amplitudes and phases depending on actual orientations. For the examples in Fig.s 1 (b) and 1 (c), these data would immediately tell the direction of arrival of the gravitation strain causing the trembling in the antenna arms (the chosen stellar pairs) as that of minimum signal. Therefore, the ultimate precision with which a direction can be recovered is sub-$mas$, 
as the sources will be in the Gaia Catalogs  \citep[see, e.g.,] []{lindegren}. Or, one could say the actual directional uncertainty is comparable to the angles separating the stellar pairs used as arms, i.e., $\sim$ 10 mas or less. Either case, these are unprecedented numbers and pinpointing source directions with such accuracies would tremendously help multi-wavelength and multi-messenger follow-up investigations.



\section{Concluding remarks and perspectives}
Immediately after the initial successful efforts at the idea of an antenna for measuring GW's using astrometry from a telescope in space  \citep[see sec. 6.4 in][]{crosta2019}, we initiated efforts to prove, with simulations and laboratory tests, the feasibility of an astrometric antenna based on the precepts described in this article \citep[and references therein]{astra1, astra2, riva20}. We refer to those studies for results and discussions on what can be currently said on implementation issues like, just to mention a few: 
i) the very possibility to build a 3-LOS multiplexing telescope \citep[and references therein]{sun}, ii) the limit of centering accuracies of star-like images on digital detectors, 
iii) actual (beyond Rayleigh's) resolution limits for the antenna arms (depending not only on  optics and detection system, but also on magnitude and color of the stellar pairs), 
iv) other natural (intrinsic or cosmic) causes of astrometric noise as stellar variability (both astrometric and photometric), and v) identify (via spatial laser metrology of critical degrees-of-freedom) and deal with instrumental noise mimicing unwanted variations of the antenna arms $\psi_{i_0,j_0}$.
In addition, we will have to simulate much more realistic scenarios (i.e., more general forms of GW's and use real-sky pairs) and conditions (realistic noise levels) to investigate viable  strategies for the actual retrieval of amplitude and phase (carrying the direction information). 
However, for these particular aspects we can certainly draw from the great amount of work done, and proven on real data by the LIGO and VIRGO collaborations.

To date,  one generally considers the global reduction process at the end of a mission like Gaia 
to extract the GW astrometric signals. 
In such a context, looking for variation of the position of a single source on the sky at nas level implies the knowledge of an "absolute" 
reference, e.g. the telescope LOS, at comparable 
precision 
$\sim 10^{-15}$. 
This is an unprecedented requirement on the reconstructed attitude 
of a science satellite. 
The proposed differential technique in \cite{astra1, astra2} aims at nas measurement over a distance of the order of 1'', thus implicitly reducing 
the relative precision requirement to $\sim 10^{-9}$, with an improvement 
of six orders of magnitude.  Also, instrument calibration requirements are strongly alleviated. 
In Gaia, we have a variation of the electro-optical response of 
hundreds of mas over the $0.5^\circ$ field, calibrated 
to the $\mu$as level. 
Assuming a linear model, the corresponding electro-optical response variation 
for an astrometric GW telescope over 1'' would be in the range of hundreds of $\mu$as, i.e. a comparable calibration "power" would be required 
to scale the measurement reliability to the nas regime. 
Actually, optimal telescope design, exploiting the higher-than-linear 
decrease of many aberrations close to the optical axis, would significantly 
reduce such instrumental contribution \citep{riva20}. 

Nevertheless, as we show in this work, the astrometric observable could amplify the GW-induced signal if one takes into account the angle between two space-like directions of light \citep{crosta2019} in the framework of general relativistic astrometric models \citep{crosta2017}. In such a case one is exempted from dealing with the satellite’s attitude and the GW astrometric measurement can be translated into an observation equation  accounting for a wide range of frequencies. 

The diversity of GW frequencies potentially included in Eq.(\ref{eqfinal}) and the appropriate modeling of the gravitational shift could improve PTA and/or LISA observations \citep{PTA, LISA} and the low frequency domain due to periodic sources (e.g., Galactic binary WDs identified by Gaia), further enhancing the mapping of the Milky Way substructures \citep{barausse}. 
Moreover, the advantage of Eq. (\ref{obseqper}) is to exploit a large number of null geodesics, so 
to better scrutinize the GW direction (a critical aspect of the GW detection). 
This same feature will also enable tests on GW polarization modes by combining different telescope orientations. \\
In conclusion, the potential of the relativistic astrometric observable advocated here for the astrometric detection and identification of gravitational waves using solely stars will push on technological advances and will significantly add to the best GW detection procedures. In fact, given that stars allow us to exploit a huge number of configurations, we could potentially monitor gravitational waves coming from any direction and, at the same time, link  the properties of a GW source with extensive statistics. Each GW mechanism has typical signatures with its own associated uncertainties that, if not instrumental, could identify the underlying physics of the corresponding GW production. This helps in a truly complementary way all of the efforts dedicated to the multiband GW searches bridging low and high frequencies at different redshifts  \citep[see, for example,][]{seddaall, highf, nanohertz}. When the two unperturbed line-of-sights, although angularly very close, are related to different stellar distances,  it would be also possible to investigate time retarded effects and test the GW speed.
In addition, suitable choice of the strain $h^{GW}$ could pave the way for new GW tests on the gravitation interaction with photons. It would suffice modeling the $F^{GW}$ function accordingly.\\

 
 \section*{Appendix A: The direction cosine for the GW observation equation}

The direction cosine of two light directions is defined as: 
 
\begin{equation}
\cos \psi_{1,2} = \frac{P(u)_{\alpha \beta} k^\alpha_1 k^\beta_2}{(\sqrt{ P(u)_{\alpha \beta} k^\alpha_1 k^\beta_1}) (\sqrt{ P(u)_{\alpha \beta} k^\alpha_2 k^\beta_2} )  }
\end{equation}

where $P(u)_{\alpha \beta}= g_{\alpha \beta } + u_\alpha u_\beta$ is the operator that projects with respect to the local barycentric observer $u^\alpha$. Then the photon 4-momentum can be decomposed as
\begin{equation}
k^{\alpha} =-(u|k)u^{\alpha} +\ell^{\alpha} \, ,
\end{equation}

where $ (u|k) = g_{\alpha \beta} u^{\alpha} k^\beta$ and  $\ell^{\alpha}$ is the spatial null vector projected on the rest space of $u^\alpha$. 
Defining 
\begin{equation}
\bar k^\alpha= -\frac{k^\alpha}{(u|k)} \, ,\,\,\,\, \bar \ell^\alpha =- \frac{\ell^{\alpha}}{(u|k)}=\bar k^\alpha -u^\alpha \,.
\end{equation}
 the direction cosine simplifies as
 \begin{equation}
\cos \psi_{1,2} = {g}_{\alpha \beta}  ({\bar \ell}^{\alpha}_1 {\bar \ell}^{\beta}_2 )_{obs}.
\label{directioncosine}
\end{equation}

The BCRS metric is defined by IAU resolutions as a post-Newtonian (pN) solution of the Einstein field equations. Thus, one has  to take into account terms of this metric accurate to the order of the GW perturbations sought for.
Dropping the sum symbol, let us express the metric including both sources as:
\begin{equation}
{g}_{\alpha \beta}=  \eta_{\alpha \beta} +  \epsilon  {h}^{SS}_{(1) \alpha \beta}+ \epsilon^2  { h}^{SS}_{ (2) \alpha \beta}+ \epsilon^3  { h}^{SS}_{ (3 )\alpha \beta}+ \epsilon^4  { h}^{SS}_{(4) \alpha \beta}+  {h}_{\alpha \beta}^{GW} + O(\epsilon^5)
\label{globalmetricapprox}
\end{equation} 
where  $\epsilon$ is of the order of $v/c$, being $v$ the typical velocity of each relevant SS metric source, and the subscripts in parenthesis indicate the order of approximation. We assume that the GW perturbations are of order $\epsilon^4$ at best, i.e., at the nanoarcsecond level.
Like for the SS metric, the SS contribution to light direction can be approximated as:
\begin{equation}
{\bar \ell}^{\alpha (SS)}_{obs} = {\bar \ell}^{\alpha}_{_0} + \epsilon  {\bar \ell}^{\alpha}_{_{(1)}} +  \epsilon^2{\bar \ell}^{\alpha}_{_{(2)}}+  \epsilon^3 {\bar \ell}^{\alpha}_{_{(3)}}+ \epsilon^4 {\bar \ell}^{\alpha}_{_{(4)}} +O( \epsilon^5).
\label{approxell}
\end{equation}

From the assumptions above, one finally finds:

\begin{equation}
\cos \psi_{1,2} =\cos \psi_{1,2}^{SS} + \eta_{\alpha \beta} ({\bar \ell}^{\alpha}_{1_0} \delta \ell_{2}^{\beta} +  {\bar \ell}^{\alpha}_{2_{0}} \delta \ell_1^{\beta})_{obs}+ h^{GW}_{\alpha \beta} {\bar \ell}^{\alpha}_{1_0} {\bar \ell}^{\beta}_{2_{0}} + O(\epsilon^5).
\label{eq:alternativy}
\end{equation}

The same can be obtained by differentiating equation (\ref{eq:alternativy}), much like the linearization procedure utilized in the Gaia mission \citep{crosta2017} to solve for the unknown astrometric parameters of each individual direction, but this time with the GW strain as unknown:

\begin{equation}
 -   \sin(\psi_{1,2}) d \psi_{1,2} =  -   \sin(\psi_{1,2}^{SS})  d \psi_{1,2}^{SS} + d F^{GW}_{1,2} +O(\epsilon^5),
\end{equation}
that, for $ \delta \psi^{GW}_{1,2}\ll 1 $ and $\psi_{1,2}= \psi_{1,2}^{SS} + \delta \psi^{GW}_{1,2}$, can be approximated as 
\begin{equation}
 -   \sin(\psi_{1,2}^{SS})( d \psi_{1,2}^{SS} + d (\delta\psi_{1,2}^{GW})) \approx   -   \sin(\psi_{1,2}^{SS})  d \psi_{1,2}^{SS} +   d F^{GW}_{1,2},
\end{equation}

or, better, as

\begin{equation}
d (\delta \psi_{1,2}^{GW}) =  - \frac{d F^{GW}_{1,2}}{\sin(\psi_{1,2}^{SS})} + O(\epsilon^5).
 \label{obseqdiff}
\end{equation}

where ${\bar \ell}^{\alpha}_{obs}$ is the local line-of-sight at the observation time.

\section*{Appendix B: The total direction cosine versus an additive GW direction cosine}

In this section we clarify the assertions made in the Introduction on the consequences, when in the presence of a passing GW,  of utilizing the direction cosine observable by simply extending what it is done in the context of the Gaia mission, where the direction cosine refers to the angle of the incoming light to the observer attitude-tetrad  $E^{\alpha}_{\hat a}$. \\

Let us consider the $\it{ a ~ priori}$  assumption that the effect of a GW is that of adding a cosine term to the direction cosine associated with the SS metric, i.e., the total (tot) direction cosine is given by:   

\begin{equation}
  \cos(\psi)^{SS} + \cos(\psi)^{GW}.
\label{totcosine}
\end{equation}
With the extra assumption that the correction $\delta \bar \ell_{(GW)}$, induced by the GW to the SS line-of-sight at the observer $\bar \ell_{(SS)}$, is (formally) known, Eq. (\ref{totcosine}) would read:
%
\begin{equation}
(\eta_{\mu \nu} + h^{SS}_{\mu \nu}) ( \bar \ell^{\mu}_{(SS)} E^{\nu}_{\hat a}) + (\eta_{\mu \nu} + h^{GW}_{\mu \nu}) (\delta\bar \ell^{\mu}_{(GW)} E^{\nu}_{\hat a}) \approx  \cos(\psi)^{SS}  + \eta_{\mu \nu } \delta \bar \ell^{\mu}_{(GW)}   E^{\nu}_{\hat a},
\label{totcosine1}
\end{equation}
implying no explicit dependence on the strain $h_{GW}$.

If, on the other hand, the GW perturbation is not a known part of the observed direction at the observer, one would have:
\begin{eqnarray}
& &  (\eta_{\mu \nu} + h^{SS}_{\mu \nu}) [ (\bar \ell^{\mu}+\delta \bar \ell^{\mu}) E^{\nu}_{\hat a}] + (\eta_{\mu \nu} +  h^{GW}_{\mu \nu}) [(\bar \ell^{\mu} + \delta\bar \ell^{\mu}) E^{\nu}_{\hat a}]\nonumber \\
&\approx&  \cos(\psi)^{SS}  + 2 \eta_{\mu \nu } \delta \bar \ell^{\mu}   E^{\nu}_{\hat a} +  \eta_{\mu \nu } \bar \ell^{\mu} E^{\nu}_{\hat a}+  h^{GW}_{\mu \nu} \bar \ell^{\mu}_{\not 0}  E^{\nu}_{\hat a},
\label{totcosine2}
\end{eqnarray}
with the flat Minkowskian metric term entering twice in the Equation for the  $\bar \ell^{\alpha}$ component. Then,  the simple addition to the "background" direction cosine in the SS metric (i.e. BCRS   would  introduce the flat Minkowskian contribution twice without a priori disentanglement from the local-line-of-sight of the GW component or, if the GW shift is considered as a separate part, it would imply to discard a priori the $h^{GW}$ strain  in the observation equation.

It is only when we drop the assumption made with Eq. (\ref{totcosine}), i.e. when working directly with the total direction 
cosine, that we finally recover an expression similar to observation equation  in the Main article as:
\begin{equation}
\cos(\psi)_{tot}=  (\eta_{\mu \nu} + h^{SS}_{\mu \nu} + h^{GW}_{\mu \nu})[ ( \bar \ell^{\mu} + \delta\bar \ell^{\mu}) E^{\nu}_{\hat a}]\approx \cos(\psi)^{SS} +  \eta_{\mu \nu}   \delta\bar \ell^{\mu}E^{\nu}_{\hat a} +  h^{GW}_{\mu \nu} \bar \ell^{\mu} E^{\nu}_{\hat a}.
\label{totcosine3}
\end{equation}
%
%
However, this form of the observation equation explicitly depends on the observer orientation, i.e., on satellite attitude, with the drawbacks mentioned in the Main article in relation to attitude errors heavily affecting the error 
budget of the measurements.

\section*{Appendix C: The astrometric gravitational shift}

The four tangent vector to a null geodesic satisfies the well known conditions:

\beq
\label{geo_gen}
k^\alpha\nabla_\alpha k^\beta=0\,, \qquad
k^\alpha k_\alpha=0\,, 
\eeq

$ \nabla_\alpha$ being the covariant derivative associated with the spacetime metric.
The decomposition of the photon 4-momentum with respect to an observer $u^{\alpha}$ implies that the trajectory is parametrized by $\sigma$ such that 

\beq
\bar k^\alpha= -\frac{k^\alpha}{(u|k)}=\frac{\rmd x^\alpha}{\rmd \sigma},
\eeq

and equation (\ref{geo_gen}) becomes
\beq
\bar k^\alpha\nabla_\alpha \bar k^\beta=-\frac{\rmd\ln{[-(u|k)]}}{\rmd\sigma}\bar k^\beta\, ,
\eeq
which is related to the affine parameter $\lambda$ by $\rmd \sigma=-(u|k) \rm d \lambda$.

It is easy to check that in the case of a static observer:
\beq
\frac{\rmd \ln{[-(u|k)] }}{\rmd\sigma}=\bar\ell^\alpha\bar\ell^\beta K_{\alpha\beta}-\bar\ell^\alpha a_\alpha
=-\bar\ell^\alpha\bar\ell^\beta \theta_{\alpha\beta}-\bar\ell^\alpha a_\alpha\,,
\eeq
where the two spatial fields coming from the splitting of the covariant derivative of $u$, i.e., $\nabla_{\beta }u^\alpha=-a^\alpha u_\beta-K^{\alpha}{}_{\beta}$, the acceleration vector $a^\alpha$ and the kinematical tensor $K^{\alpha}{}_{\beta}=\omega^{\alpha}{}_{\beta} -\theta^{\alpha}{}_{\beta}$ are expressed as a combination of the vorticity and expansions of the congruence of curves related to fiducial observers $u^{\alpha}$. 
Thus, the geodesic equation transforms into 
\beq
\frac{\rmd \bar k^\alpha}{\rmd \sigma}+ \Gamma^\alpha{}_{\mu\nu}\bar k^\mu\bar k^\nu
-\left[\bar\ell^\mu\bar\ell^\nu \theta_{\mu\nu}+\bar\ell^\mu a_\mu\right]\bar k^\alpha=0\,,
\eeq
or 
\begin{eqnarray}
\label{eq:geodfull}
\frac{\rmd \bar\ell^\alpha}{\rmd \sigma}&+&
\Gamma^\alpha{}_{\mu\nu}\bar\ell^\mu(\bar\ell^\nu+u^\nu)+a^\alpha-K^\alpha{}_\sigma \bar\ell^\sigma 
-\left[\bar\ell^\mu\bar\ell^\nu \theta_{\mu\nu}+\bar\ell^\mu a_\mu\right](\bar\ell^\alpha+u^\alpha)=0\, ,
\end{eqnarray}
for the  unknown local line-of-sight $\bar \ell^\alpha$.
In case of static observers, the kinematical fields reduce to
\beq
a^i=\partial_0 h_{0i}-\frac12\partial_i h_{00}
=h_{0i,0}-\frac12 h_{00,i}\,,\qquad
\theta_{ij}=\frac12 h_{ij,0}\,,\qquad
\omega_{ij}
=-h_{0[i,j]}\,.
\eeq

Note that considering the metric as an approximate solution of the Einstein field equation composed of a background part plus a GW perturbation, also the affine coefficients can be split respectively into two parts:
\beq
 \Gamma^\alpha{}_{\mu\nu} = \frac12 g^{\alpha\rho}\left(g_{\rho\beta,\gamma}+g_{\rho\gamma,\beta}-g_{\beta \gamma,\rho}\right)  \approx \Gamma^{\alpha (SS)}_{\beta \gamma} + \Gamma^{\alpha (GW)}_{\beta \gamma}\,, 
 \label{eq:gammafull}
\eeq
 at the first order of the perturbation.
The same for the parameter sigma:
\beq 
d\sigma = - (u|k) d\lambda =  - [ (g_{\alpha \beta}^{(SS)} + h_{\alpha \beta}^{(GW)}) \, u^{\alpha} k^{\beta} ]d\lambda. 
\label{eq:paraffine}
\eeq
The TT gauge choice implies that the second term of equation (\ref{eq:paraffine}) does not contribute.

All of the above implies the possibility again to split equation (\ref{eq:geodfull}) into the Solar System part plus the GW one, thus it allows to integrate separately each  term . As a matter of fact, at the order of $\epsilon^4$, it is possible to isolate from equation (\ref{eq:geodfull}) the contribution of the GW part and obtain
\beq
 \frac{d \delta \ell^i}{d \sigma}  \approx -  \frac{1}{2} \bar\ell^j_{_0} \bar\ell^k_{_0} (2 h^{GW}_{ij,k} -  h^{GW}_{jk,i}) -  \bar\ell^j_{_0}  h^{GW}_{ij,0}  + \frac{\bar\ell^j_{_0} \bar\ell^k_{_0} \bar\ell^i_{_0}}{2}  h^{GW}_{jk,0}.
 \label{eq:geoGW}
\eeq
Since $d\sigma = d\lambda + O(\epsilon^2)$ and assuming for the photon trajectory $x^0(\sigma) = x^0_{obs} + \sigma + O(\epsilon^2)$ and   $x^i(\sigma) = x^i_{obs}+ \ell^i_{ _0} \sigma + O(\epsilon^2)$, the argument of $h_{ij}$ becomes:
\beq
\tilde k_{\alpha} x^{\alpha} = \tilde k_0 \sigma (1 +  p \cdot \bar\ell_{_0}) + \tilde \psi + O(\epsilon^2)
\eeq
where $\tilde \psi =  \tilde k_{\alpha} x_{obs}^{\alpha}$ can be considered a phase term. Then,
\beq
h_{ij,0} = \frac{1}{(1 +  p \cdot \bar\ell_{ _0} )} \frac{dh} {d \sigma}, \,\,\,\, 
h_{ij,k} = \frac{ p^k}{ (1 + p \cdot \bar\ell_{ _0}) } \frac{dh} {d \sigma}.
\eeq
Via a direct integration of equation (\ref{eq:geoGW}) we easily obtain the gravitational shift of the local direction:
\beq
\int^{\sigma_*}_{\sigma_{obs}}  d \delta \ell^i =  \frac{\bar \ell^i_{_0} + p^i}{2 (1 +  p \cdot \bar\ell_{_0})} \bar \ell^j_{_0} \bar \ell^k_{_0} \int^{\sigma_*}_{\sigma_{obs}}   d h_{jk}^{(GW)}  (\sigma)  -   \frac{\bar \ell^j_{_0}} {2}   \int^{\sigma_*}_{\sigma_{obs}} d h_{ij}^{(GW)}  (\sigma) + O(\epsilon^5),
\eeq
which coincides  with the result in  \citet{book} when the distance to the stellar source  is many gravitational waves away, namely the detection occurs in the far-away wave zone.
{}

\end{document}